\documentclass[conference, a4paper, 10pt, twocolumn]{IEEEtran}

\usepackage[utf8]{luainputenc}

\usepackage[noadjust]{cite}

\usepackage{graphicx}

\usepackage{pgfplots}
\newlength\figureheight
\newlength\figurewidth 
\pgfplotsset{compat=newest}
\pgfplotsset{plot coordinates/math parser=false} 
\pgfplotsset{every x tick label/.append style={font=\scriptsize, yshift=0ex}}
\pgfplotsset{every y tick label/.append style={font=\scriptsize, xshift=0ex}}
\pgfplotsset{every axis legend/.append style={font=\scriptsize}}

\usepackage{units}

\usepackage[cmex10]{amsmath}

\usepackage{amssymb,mathtools}

\usepackage{array}

\usepackage[caption=false,font=footnotesize]{subfig}

\usepackage{url}

\usepackage{minamakron}
\let\oldmathbf\mathbf

\renewcommand{\mathbf}[1]{{\boldsymbol{\oldmathbf{#1}}}}

\renewcommand{\url}[1]{#1}

\newcommand{\utannotation}[1]{}

\newcommand*\circled[1]{\tikz[baseline=(char.base)]{
		\node[shape=circle,draw,inner sep=0.1ex] (char) {#1};}}

\newcommand{\jia}{\normalfont\scriptsize \circled{1}}
\newcommand{\yi}{\normalfont\scriptsize \circled{2}}
\newcommand{\bing}{\normalfont\scriptsize \circled{3}}

\usepackage[cal=boondoxo]{mathalfa}
\begin{document}

	\title{Out-of-Band Radiation Measure for MIMO Arrays with Beamformed Transmission}

	\author{\IEEEauthorblockN{Christopher Mollén\hspace{0.1em}$^{\text{\jia}}$, Ulf Gustavsson\hspace{0.1em}$^{\text{\yi}}$, Thomas Eriksson\hspace{0.1em}$^{\text{\bing}}$, Erik G. Larsson\hspace{0.1em}$^{\text{\jia}}$%
		}
		\IEEEauthorblockA{\llap{\raisebox{\depth}{\scriptsize${\text{\jia}}$}}\enskip Linköping University, Dept.\ of Electrical Engineering, 581 83 Linköping, Sweden}
			\IEEEauthorblockA{\llap{\raisebox{\depth}{\scriptsize${\text{\yi}}$}}\enskip Ericsson Research, Lindholmspiren 11, 417 56 Gothenburg, Sweden}
			\IEEEauthorblockA{\llap{\raisebox{\depth}{\scriptsize${\text{\bing}}$}}\enskip Chalmers University of Technology, Dept.\ of Signals and Systems, 412 96 Gothenburg, Sweden}
		\thanks{The research leading to these results has received funding from the European Union Seventh Framework Programme under grant agreement number \textsc{ict}-619086 (\textsc{mammoet}) and the Swedish Research Council (Vetenskapsrådet).}
	}

	\maketitle

	\begin{abstract}
		The spatial characteristics of the out-of-band radiation that a multiuser \MIMO system emits in the environment, due to its power amplifiers (modeled by a polynomial model) are nonlinear, is studied by deriving an analytical expression for the continuous-time cross-correlation of the transmit signals.  At a random spatial point, the same power is received at any frequency on average with a \MIMO base station as with a \textsc{siso} base station when the two radiate the same amount of power.  For a specific channel realization however, the received power depends on the channel.  We show that the power received out-of-band only deviates little from the average in a \MIMO system with multiple users and that the deviation can be significant with only one user.  Using an ergodicity argument, we conclude that out-of-band radiation is less of a problem in massive \MIMO, where total radiated power is lower compared to \textsc{siso} systems and that requirements on spectral regrowth can be relaxed in \MIMO systems without causing more total out-of-band radiation.
	\end{abstract}


	\section{Introduction}
	\IEEEPARstart{O}{ut-of-band} radiation is the undesired power of a signal at frequencies outside the allocated frequency band.  Such power usually arises from nonlinear circuits and can potentially disturb concurrent transmission in adjacent bands.  Therefore, many standards, e.g. \textsc{lte} \cite{3GPP_TS36.141_LTE_BS_testing}, limit the amount of out-of-band radiation that is allowed to be emitted.  Traditionally, out-of-band radiation has been measured on a per-antenna basis.  In a \MIMO setting, where many antennas concurrently transmit, this is not necessarily a sensible way to measure.  The radiated power from the transmitting antennas builds up constructively or destructively in the air and the amount of out-of-band radiation that disturbs transmission in adjacent bands can thus be greater or smaller than what was emitted from any single antenna.  Not to disturb other communication, the out-of-band radiation should therefore instead be limited on the basis of what is actually received by the users of adjacent bands.  
	
	In this article, we study the spatial distribution of the out-of-band radiation in order to gain some fundamental insight into its behavior in multi-antenna systems with nonlinear amplifiers, and to understand how it should be appropriately measured.  This will be an important aid for the standardization process of future communication systems, which are envisioned to incorporate base stations with hundreds or thousands of antennas---so called \emph{massive \MIMO}---to increase spectral efficiency and radiated energy efficiency by orders of magnitude compared to systems used today \cite{6690}. 
	
	We use a polynomial model to characterize the nonlinear power amplifier of the base station and to derive an analytical expression for the cross-correlation matrix of the downlink transmit signals in a \MIMO system. Using the cross-correlation matrix, the spatial distribution of the power received at different frequencies can be computed and analyzed.  Just like other sources of uncorrelated interference tend to become negligible at the user-side in massive \MIMO because of the big array gain of the system \cite{bjornson2013massive, athley2015analysis}, we find that the power received outside the band is substantially lower when the number of antennas is big than when only a single antenna is used.  This means that the linearity requirement of the base station hardware can be lowered in massive \MIMO relative to single-antenna systems.  Further, we define a measure of out-of-band radiation for \MIMO systems based on over-the-air received powers and show that it is substantially the same as measuring out-of-band radiation on a per-antenna basis when the small-scale fading coefficients to the different antennas are uncorrelated.  Out-of-band radiation can therefore be measured on a per-antenna basis also in \MIMO.
	
	The phenomenon of out-of-band radiation in single-antenna systems has been thoroughly studied before, see for example \cite{gard99microwave}.  Methods developed to mitigate out-of-band radiation, such as digital pre-distortion, are also well known \cite{kim2001digital}.  Many of these methods are, however, undesirable in a massive \MIMO system due to the great number of radio chains.  Empirical studies of out-of-band radiation have been done in massive \MIMO systems, see \cite{UGUSGC14}.  To the authors' knowledge, however, the spatial distribution of out-of-band radiation from a multi-antenna base station, which is the subject of this study, has not been analyzed before. 
	
	\section{Notation}
	The elementwise complex conjugate of the matrix $\mathbf{M}$ is denoted $\mathbf{M}^*$, its Hermitian transpose $\mathbf{M}^\conjtr$ and its transpose $\mathbf{M}^\tr$.  If $\mathbf{M}$ is a Hermitian matrix, $\lambda_\text{max}(\mathbf{M})$ denotes its principal eigenvalue.  If $\mathbf{a}(t) \triangleq (a_1(t), a_2(t), \ldots)^\tr$ and $\mathbf{b}(t) \triangleq (b_1(t), b_2(t), \ldots)^\tr$, where $\{a_i(t), b_i(t)\}$ are jointly weakly stationary random processes, their cross-correlation function is denoted by $\mathbf{R}_{\mathbf{ab}}(\tau) \triangleq \mathsf{E}\big[\mathbf{a}^*(t) \mathbf{b}^\tr(t+\tau)\big]$.  Similarly, the autocorrelation function of a discrete-time signal 	$\mathbf{a}[n] \triangleq (a_1[n], \ldots, a_2[n])^\tr$ is denoted $\mathbf{R}_{\mathbf{aa}}[\nu] \triangleq \Expectation{\mathbf{a}^*[n] \mathbf{a}^\tr[n+\nu]}$.  Furthermore, $\big(\phi(\tau) \star \psi(\tau)\big)(t)$ denotes the convolution at $t$ between the functions $\phi(\tau)$ and $\psi(\tau)$, $\delta(\tau)$ the continuous-time Dirac distribution and $\mathbf{I}_K$, $\mathbf{0}_K$ the $K$\mbox{-}dimensional identity and all-zero matrices.
	
\section{Downlink System Model}	
	The base station transmits the digital signals $\mathbf{x}[n] \triangleq (x_1[n], \ldots, x_M[n])^\tr$ on its $M$ antennas by pulse-amplitude modulating them with the pulse $p(\tau)$ into the analog signal
	\begin{align}
	\mathbf{x}(t) \triangleq \begin{pmatrix}
	x_1(t)\\\vdots\\x_M(t)
	\end{pmatrix} = \smash{\sum_{n}} \mathbf{x}[n] p(t - nT + \Psi),
	\end{align}
	where $T$ is the symbol duration and $\Psi$ is a random variable\footnote{The introduction of $\Psi$ is a way to make pulse-amplitude modulation preserve stationarity \cite{papoulis2002probability}; it only appears in this equation.  } that is uniformly distributed on the interval $0 \leq \Psi < T$.  The bandwidth of the pulse $p(\tau)$ is assumed to be equal to the bandwidth $B$ that is allocated to the base station.  The signal $\mathbf{x}(t)$ is amplified to transmit power into $\mathbf{y}(t) \triangleq (y_1(t), \ldots, y_M(t))^\tr$, where the amplification is modeled as
	\begin{align}\label{eq:PA_IO_relation}
	y_m(t) = \sum_{p=1}^{P} \int\limits_{-\infty}^{\infty} b_{mp}(t-\tau) x_m(\tau) |x_m(\tau)|^{2(p-1)} \mathrm{d}\tau,
	\end{align}
	where $b_{mp}(\tau)$ is the impulse response of the nonlinear $p$\mbox{-}th order term of the $m$\mbox{-}th amplifier \cite{kim2001digital}.  Note that this polynomial model is a special case of the more general Volterra series \cite{Schetzen80}: all kernels outside the diagonal are set to zero and all dynamic memory is removed.  
		
	We now let $r_{\mathbf{\theta}}(t)$ denote the received signal at a point $\mathbf{\theta}$ in space.  The received signal can then be computed as
	\begin{align}
		r_{\mathbf{\theta}}(t) = \sqrt{\beta_\mathbf{\theta}} \int\limits_{-\infty}^{\infty} \mathbf{h}_{\mathbf{\theta}}^\tr(\tau) \mathbf{y}(t-\tau) \mathrm{d}\tau,
	\end{align}
	where $\mathbf{h}_{\mathbf{\theta}}(\tau)$ is the impulse response of the small-scale fading from the array to the point $\mathbf{\theta}$ and $\beta_\mathbf{\theta} \in \mathbb{R}^+$ a large-scale fading coefficient, which models signal attenuation due to both distance and shadowing.
	
	\section{Base Station Radiation Pattern}
	We assume that the base station is serving $K$ single-antenna users and that the $M$ transmit signals are produced by linear precoding as:
	\begin{align}\label{eq:precoding}
		\mathbf{x}[n] = \sum_{\ell} \mathbf{W}[\ell] \mathbf{D}_\mathbf{\xi}^{1/2} \mathbf{s}[n-\ell],
	\end{align}
	where $\mathbf{s}[n] \triangleq (s_1[n], \ldots, s_K[n])^\tr$, $s_k[n]$ is the symbol to be transmitted to user $k$ at symbol time $n$, $\mathbf{D}_\mathbf{\xi} \triangleq \operatorname{diag}(\mathbf{\xi})$ is a diagonal matrix with the relative power allocations $\mathbf{\xi} \triangleq (\xi_1,\ldots,\xi_K)^\tr$, for which $\xi_k \in \mathbb{R}^+$ and $\sum_{k=1}^K \xi_k = 1$, on its diagonal and $\{\mathbf{W}[\ell]\}$ is the impulse response of the precoder.  
	
	The discrete-time channel is given by
	\begin{align}
	\mathbf{H}[\ell] \triangleq \Big(p(\tau) \star \mathbf{H}(\tau) \star p^*(-\tau)\Big)(\ell T),
	\end{align}
	where $\mathbf{H}(\tau) \triangleq (\mathbf{h}_{\mathbf{\theta}_1}(\tau), \ldots, \mathbf{h}_{\mathbf{\theta}_K}(\tau))^\tr$ and $\mathbf{\theta}_k$ is the location of user $k$.  The simplest linear precoder is the maximum-ratio precoder, whose impulse response is given by $\mathbf{W}[\ell] = \alpha \mathbf{H}^\conjtr[-\ell]$, where $\alpha$ is a real-valued normalization factor that is chosen such that $\sum_{\ell}\|\mathbf{W}[\ell]\|^2_\mathsf{F} = K$.  Other common precoders are zero-forcing precoding and regularized zero-forcing precoding, see e.g.\ \cite{6832894,mollen2015waveforms}.  We assume that the base station knows $\mathbf{H}[\ell]$ perfectly. 
	
	Further, we assume that $\mathbf{s}[n]$ is a circularly symmetric i.i.d.\ stationary process, for which
	\begin{align}
		\mathbf{R}_{\mathbf{ss}}[\nu] \utannotation{\triangleq \Expectation{\mathbf{s}^*[n] \mathbf{s}^\conjtr[n+\nu]}} = \begin{cases}
		\mathbf{I}_{K}, \quad & \text{if } \nu = 0\\
		\mathbf{0}_{K}, & \text{otherwise}
		\end{cases}.
	\end{align}
	Because of the multiuser precoding in \eqref{eq:precoding} and of the central limit theorem, the distribution of the discrete-time transmit signals $\mathbf{x}[n]$ is close to circularly symmetric Gaussian.  Note that this is true independently of whether \OFDM or single-carrier transmission is used and independently of the order of the symbol constellation \cite{mollen2015waveforms}.  The autocorrelation function of the unamplified transmit signals $\mathbf{x}[n]$ in a given coherence interval (the expectation is taken with respect to the symbols conditioned on the small-scale fading) is
	\begin{align}
	\null\hskip-0.3em& \mathbf{R}_{\mathbf{xx}}[\nu] = \notag\\
		 \null\hskip-0.3em& \Expectation{\hskip-0.4em\Big(\!\sum_{\ell}\! \mathbf{W}^*[\ell] \mathbf{D}_\mathbf{\xi}^{1/2} \mathbf{s}^*[n{-}\ell]\!\Big)\!\! \Big(\!\sum_{\ell'}\! \mathbf{s}^\tr[n{+}\nu{-}\ell'] \mathbf{D}_\mathbf{\xi}^{1/2} \mathbf{W}^\tr\![\ell']\!\Big)\hskip-0.4em}\notag\\
		\null\hskip-0.3em&=\smash{\sum_{\ell}}\rule[-2ex]{0pt}{3ex}\mathbf{W}^*[\ell] \mathbf{D}_\mathbf{\xi} \mathbf{W}^\tr[\nu{+}\ell].
	\end{align}
	For example, if maximum-ratio precoding is done, $\mathbf{R}_{\mathbf{xx}}[\nu] = \alpha^2 \sum_{\ell} \mathbf{H}^\tr[\ell] \mathbf{D}_\mathbf{\xi} \mathbf{H}^*[\ell - \nu]$.  The pulse-amplitude modulated $\mathbf{x}(t)$ thus has the autocorrelation function
	\begin{align}
		\mathbf{R}_{\mathbf{xx}}(\tau) = \frac{1}{T}\sum_{\nu=-\infty}^{\infty} \mathbf{R}_{\mathbf{xx}}[\nu] \big(p(t) \star p^*(-t)\big)(\tau - \nu T).
	\end{align}
	
	The cross-correlation of the transmit signal is thus:
	\begin{align}
		R_{y_my_{m'}}(\tau) = \mathsf{E}\Big[\sum_{p=1}^{P} \int\limits_{-\infty}^{\infty} b_{mp}^*(t-\lambda) x_m^*(\lambda) |x_m(\lambda)|^{2(p-1)}\mathrm{d}\lambda\notag\\[-1.5ex]
		\sum_{p'=1}^{P} \int\limits_{-\infty}^{\infty} b_{m'p'}(t + \tau -\lambda') x_{m'}(\lambda') |x_{m'}(\lambda')|^{2(p'-1)} \mathrm{d}\lambda'\Big]\\
		= \sum_{p=1}^{P} \sum_{p'=1}^{P} \int\limits_{-\infty}^{\infty} \int\limits_{-\infty}^{\infty} b_{mp}^*(t-\lambda) b_{m'p'}(t+\tau-\lambda')\notag\\
		\smash{\underbrace{\Expectation{x_m^*(\lambda)x_{m'}(\lambda')|x_{m}(\lambda)|^{2(p-1)}|x_{m'}(\lambda')|^{2(p'-1)}}}_{\triangleq \xi_{mm'}^{(p,p')}(\lambda,\lambda')}}\mathrm{d}\lambda\mathrm{d}\lambda' \label{eq:tx_cross_corr}
	\end{align} \vspace{1ex}
	
	\noindent In the last step, the variable $t$ just translates the integrand.  The integral thus does not depend on $t$ and the transmit signals are therefore weak-sense stationary.  Because odd moments of Gaussian random variable are zero, we see that $\xi_{mm'}^{(p,p')}(\lambda,\lambda')$ is zero for $m \neq m'$, for all $p$, $p'$, $\lambda$, $\lambda'$, if the unamplified signals $x_m(t)$ are uncorrelated across the antennas.  This means that, when $\mathbf{R}_\mathbf{xx}(\tau)$ is diagonal, $\mathbf{R}_\mathbf{yy}(\tau)$ is diagonal too.
	
	Using the moment theorem for Gaussian random variables \cite{reed1962moment}, $\xi_{mm'}^{(p,p')}(\lambda,\lambda')$ can be computed for any $m,m',p,p'$, e.g.,
	\begin{flalign}
		&\xi_{mm'}^{(1,1)}(\lambda,\lambda') = R_{x_mx_{m'}}(\lambda' - \lambda)&&\\
		&\xi_{mm'}^{(1,2)}(\lambda,\lambda') = 2\sigma_{x_m}^2 R_{x_mx_{m'}}(\lambda' - \lambda)&&\\
		&\xi_{mm'}^{(2,2)}\hskip-0.2em(\lambda,\lambda') {=} 2R_{x_mx_{m'}}\hskip-0.2em(\lambda' \! {-} \lambda) \!\Big(\!2\sigma_{x_m}^2 \! \sigma_{x_{m'}}^2 \hskip-0.2em{+} \big|R_{x_mx_{m'}}\hskip-0.2em(\lambda' \! {-} \lambda)\big|^2\Big),&&
	\end{flalign}
	where $\sigma_{x_m}^2 \triangleq R_{x_mx_m}(0)$.  Furthermore, we note that 
	\begin{align}
		\xi_{mm'}^{(p,p')}(\lambda,\lambda') = {\xi_{m'm}^{(p',p)}}^{\hskip-0.1em*}\hskip-0.3em(\lambda',\lambda).
	\end{align}
	
	To study the radiation pattern of the array at different frequencies, we define the frequency response of the channel to the point $\mathbf{\theta}$ as:
	\begin{align}
		\mathbf{\tilde{h}}_\mathbf{\theta}(f) \triangleq \int\limits_{-\infty}^\infty \rule{0pt}{4ex}\mathbf{h}_\mathbf{\theta}(\tau) e^{-j2 \pi \tau f} \mathrm{d}\tau.
	\end{align}
	Let $\mathbf{R}_{\mathbf{yy}}(\tau)$ be the matrix, whose $(m,m')$\mbox{-}th element is $R_{y_my_{m'}}(\tau)$.  The radiation pattern is given by the power spectral density
	\begin{align}
		\mathbf{S}_{\mathbf{yy}}(f) \triangleq \int\limits_{-\infty}^{\infty} \mathbf{R}_{\mathbf{yy}}(\tau) e^{-j 2 \pi \tau f}\mathrm{d}\tau
	\end{align}
	and the power received at the point $\mathbf{\theta}$ at frequency $f$ is
	\begin{align}\label{eq:rx_PSD}
		S_{\mathbf{\theta}}(f) \triangleq \beta_\mathbf{\theta} \mathbf{\tilde{h}}_{\mathbf{\theta}}^\conjtr(f) \mathbf{S}_{\mathbf{yy}}(f) \mathbf{\tilde{h}}_{\mathbf{\theta}}(f).
	\end{align}
	Note that the power radiated by the base station at frequency $f$ is 
	\begin{align}
		S_\text{tx}(f) \triangleq \trace(\mathbf{S}_\mathbf{yy}(f))
	\end{align}
	and that the average received power at a point, where $\mathbf{\tilde{h}}_\mathbf{\theta}(f)$ is independent of $\mathbf{S}_\mathbf{yy}(f)$ and the fading at the different antennas are zero mean and uncorrelated $\mathsf{E}\big[\mathbf{\tilde{h}}_\mathbf{\theta}(f) \mathbf{\tilde{h}}\llap{\phantom{h}}_\mathbf{\theta}^\conjtr(f) \big] = \mathbf{I}_M$, is
	\begin{align}\label{eq:average_rx_power}
		\Expectation{S_\mathbf{\theta}(f)} = \beta_\mathbf{\theta} \Expectation{S_\text{tx}(f)}.
	\end{align}
	The expectation is over all small-scale fading, also over the channels to the users, on which the precoding is based.
	
	\section{Measures of Out-of-Band Radiation}
	To constrain the amount of out-of-band radiation a base station radiates, it is important to be able to easily measure it at the base station.  In this section, we study the measure conventionally used in single-antenna systems and generalize it to multi-antenna systems.  We also propose a framework to analyze how the transmitted signal is beamformed at different frequencies---in-band and out-of-band.
	
	\subsection{The Traditional Single-Antenna Setting}
	Traditionally, the \emph{transmitted} out-of-band radiation has been measured at the antenna port in terms of the Adjacent-Channel Leakage Ratio (\ACLR).  Let $S_{yy}(f)$ be the power spectral density of the transmit signal in a single-antenna base station.  Then \ACLR is defined as \cite{3GPP_TS36.141_LTE_BS_testing, schreurs2009rf}:
\begin{align}\label{eq:ACLR}
	\ACLR \triangleq \frac{\max\!\big\{\!\int_{-3B/2}^{-B/2} S_{yy}(f) \mathrm{d}f, \int_{B/2}^{3B/2} S_{yy}(f) \mathrm{d}f\big\}}{\int_{-B/2}^{B/2} S_{yy}(f)\mathrm{d}f}.
\end{align}
The measure compares the amount of power that has leaked over to an immediately adjacent band, which is assumed to have the same width $B$ as the allocated band, to the power in the allocated band.  The first term in the numerator of \eqref{eq:ACLR} is the power in the band just to the left of the allocated band and the second term that in the band to the right.  

We let $\mathbf{\tilde{h}}_\mathbf{\theta}(f) = \tilde{h}_\mathbf{\theta}(f)$ be the frequency response from the single-antenna base station to the point $\mathbf{\theta}$.  If the antenna gain is constant over the frequency band $[-3B/2, 3B/2]$, then \ACLR equivalently can be measured in a fading environment in the air too as
\begin{align}\label{eq:airACLR}
	\ACLR = \frac{\max\{\int_{-3B/2}^{-B/2} \Expectation{S_\mathbf{\theta}(f)} \mathrm{d}f, \int_{B/2}^{3B/2}\Expectation{S_\mathbf{\theta}(f)} \mathrm{d}f\}}{\int_{-B/2}^{B/2} \Expectation{S_\mathbf{\theta}(f)}\mathrm{d}f},
\end{align}
where averaging is done over the small-scale fading.  Note that, because of the averaging, this ratio is the same at every location $\mathbf{\theta}$ and is equal to $\ACLR$ in \eqref{eq:ACLR}.  A fading environment can be artificially created in a reverberation chamber, which would lend itself to practical measurements of this kind \cite{holloway2006use}.

\subsection{The Multi-Antenna Setting}
The most straightforward way to generalize the \ACLR measure to a multi-antenna setting is to define a per-antenna \ACLR as
	\begin{align}
	&\ACLR_m \triangleq \notag\\
		&\frac{\max\!\big\{\!\int_{-3B/2}^{-B/2} \Expectation{S_{y_my_m}\!(f)} \mathrm{d}f, \int_{B/2}^{3B/2} \Expectation{S_{y_my_m}\!(f)} \mathrm{d}f\big\}}{\int_{-B/2}^{B/2} \Expectation{S_{y_my_m}\!(f)} \mathrm{d}f}.\label{eq:perantenna_ACLR}
	\end{align}
	Since signals from a multi-antenna base station combine in the air however, there is a chance that the received power in an adjacent band is different from the transmitted power.  Therefore it remains to determine what the per-antenna \ACLR says about how much a victim receiver, who operates in an adjacent band, really is disturbed.
	
	Based on the observation in \eqref{eq:airACLR}, we define a measure that generalizes the \ACLR concept to multi-antenna transmission.  We define the \textsc{mimo-aclr} as
	\begin{align}
		&\hspace{-1em}\forkortning{MIMO\text{-}ACLR}(\mathbf{\theta}) \triangleq \notag\\
		&\frac{\max\{\int_{-3B/2}^{-B/2} \Expectation{S_\mathbf{\theta}(f)} \mathrm{d}f, \int_{B/2}^{3B/2}\Expectation{S_\mathbf{\theta}(f)} \mathrm{d}f\}}{\int_{-B/2}^{B/2} \Expectation{S_\mathbf{\theta}(f)}\mathrm{d}f}.\label{eq:mimoaclr}
	\end{align}
	In the definition, the expectation is taken with respect to the small-scale fading.  The small-scale fading $\mathbf{\tilde{h}}_\mathbf{\theta}(f)$ is assumed to be independent of that of the users $\mathbf{\tilde{h}}_{\mathbf{\theta}_k}(f)$, for all $k$, so that $\mathbf{\tilde{h}}_\mathbf{\theta}(f)$ and $\mathbf{S_\mathbf{yy}}(f)$ are independent.
	
	We show that the measure \forkortning{mimo-aclr} has the following properties, if $\mathsf{E}\big[\mathbf{\tilde{h}}_\mathbf{\theta}(f) \mathbf{\tilde{h}}\llap{\phantom{h}}_\mathbf{\theta}^\conjtr(f) \big] = \mathbf{I}_M$, for all $\mathbf{\theta}$:
	\begin{description}
		\item[P1] It does not depend on the large-scale fading $\beta_\mathbf{\theta}$ and is the same for all $\mathbf{\theta}$.
		
		\item[P2] It does not change if the transmitted signal is scaled.
		
		\item[P3] It is equal to the per-antenna $\ACLR_m$ and to the $\ACLR$ of a single-antenna system with the same radiated power.\label{prop:MIMOACLR_ACLR_same}
	\end{description}
	The properties P1, P2 and P3 follow from \eqref{eq:average_rx_power}, which gives
	\begin{align}
	&\hspace{-1em}\MIMOACLR = \notag\\
		 &\frac{\max\big\{\int_{-3B/2}^{-B/2} \Expectation{S_\text{tx}(f)} \mathrm{d}f, \int_{B/2}^{3B/2} \Expectation{S_\text{tx}(f)} \mathrm{d}f\big\}}{\int_{-B/2}^{B/2} \Expectation{S_\text{tx}(f)} \mathrm{d}f},
	\end{align}
	where the argument $\mathbf{\theta}$ has been dropped.  
	
	Further, we conjecture that the measure \MIMOACLR has this property:
	\begin{description}
		\item[C1] It depends only weakly on the power allocations $\{\xi_k\}$ and the path losses $\{\beta_{\mathbf{\theta}_k}\}$ of the users.
	\end{description}
	The conjectured property C1 remains a conjecture in this study.  It is however made plausible by the fact that the optimal transmit direction of each user $k$ does not depend on its path loss $\beta_{\mathbf{\theta}_k}$ in massive \MIMO, see \cite{Sanguinetti2014Optimal}.  
	
	It is important to note that, due to its high array gain, a massive \MIMO system can radiate less power than a single-antenna system for a given performance requirement.  Hence, even if the $\ACLR$ in a single-antenna system and the $\MIMOACLR$ in a massive \MIMO system are the same, the absolute amount of interfering power a victim that operates in an adjacent band suffers from is lower in the massive \MIMO system than in the single-antenna system.  Property P3 of the \MIMOACLR measure thus suggests that the \MIMOACLR for massive \MIMO can be higher than \ACLR can be for a single-antenna system without disturbing communication in adjacent bands more---the difference between \MIMOACLR and \ACLR roughly being equal to the array gain of the massive \MIMO system.
	
	\subsection{Worst-Case Out-of-Band Radiation}
	If coding can be done over multiple coherence intervals, then only the average amount of received out-of-band radiation is relevant for a victim.  However, there are cases, where the channels are correlated, e.g.\ if a victim follows the movement of a served user, or where coding cannot be done over multiple coherence intervals, e.g.\ because of latency constraints or because the fading is static as in a line-of-sight scenario.  In these cases, one has to study whether there are points, to which the out-of-band radiation is beamformed, in order to protect victims in every coherence interval.  To study whether there are such points, we study the maximum power spectral density, which is defined as
	\begin{align}
		S_\text{max}(f) \triangleq \lambda_\text{max}\big( \mathbf{S}_{\mathbf{yy}}(f) \big).
	\end{align}
	This corresponds to the highest normalized power received at a given frequency at any point, i.e.
	\begin{align}
		S_\text{max}(f) \beta_\mathbf{\theta} \|\mathbf{\tilde{h}}_\mathbf{\theta}(f)\|^2 \geq S_\mathbf{\theta}(f), \quad\forall\mathbf{\theta}.
	\end{align}
	
	Note that $S_\text{max}(f)$ bounds the \emph{maximum} received power density at frequency $f$ for all channel vectors $\mathbf{\tilde{h}}_\mathbf{\theta}(f)$.   There is a possibility, however, that the maximizing channel vector has zero probability to show up in the physical environment.  The measure might therefore be a rather loose upper bound, in the sense that the maximum adjacent-band power it indicates is rarely seen by a victim user.
	
\section{Numerical Examples}\label{sec:numerical_analysis}
	In this section, the spatial distribution of the out-of-band radiation is studied for some representative scenarios.  All continuous-time signals are simulated with $\kappa = 5$\mbox{-}times oversampling.  The code to reproduce the plots can be found at \url{https://github.com/OOBRadMIMO/NumericalResults}.
	
	\subsection{Assumptions of the Numerical Analysis}
	A memory-less, third order polynomial model is assumed, where $b_{mp}(\tau) = b_{mp} \delta(\tau)$, for $p = 1,2, \forall m$, and $b_{mp}(\tau) = 0$, for $p > 2$.  Then the cross-correlation in \eqref{eq:tx_cross_corr} simplifies into
	\begin{multline}
	R_{y_my_{m'}}(\tau) = b_{m1}^* b_{m'1} R_{x_mx_{m'}}(\tau) + 2 R_{x_mx_{m'}}(\tau) \\
	\times \big( b_{m1}^* b_{m'2} \sigma_{x_m}^2 + b_{m2}^* b_{m'1} \sigma_{x_{m'}}^2 \\
	+ b_{m2}^* b_{m'2} (2\sigma_{x_m}^2 \sigma_{x_{m'}}^2 + |R_{x_mx_{m'}}(\tau)|^2) \big).
	\end{multline}
	
	We set $b_{m1} = 1$ and $b_{m2} = -0.03491 + j0.005650$ (extracted through linear regression from measurements on the class~\textsc{ab} amplifier that can be run from \cite{chalmersWeblab}), for all $m$, and made the amplifier operate at its \unit[1]{dB}\mbox{-}compression point.  As pulse shaping filter, we chose a root-raised cosine with roll-off 0.22, as in \forkortning{lte} \cite{3GPP_TS36.141_LTE_BS_testing}, which gives the normalized bandwidth $BT = 1.22$.
	
	Two channel scenarios are considered: line-of-sight and independent Rayleigh fading channels.  For simplicity, all users are assumed to be on the same distance from the base station and experience the same large-scale fading, i.e. $\beta_{\mathbf{\theta}_k} = 1$ for all $k$.  Equal power allocation is applied, i.e.\ $\xi_k = 1/K$ for all $k$.  

	In the studied line-of-sight scenario there is only one path between each antenna and each user: the direct non-obscured path.  Furthermore, a uniform linear array is considered.  Denote the angle to the $k$\mbox{-}th user by $\theta_k$.  The channel to user $k$ is then given by 
	\begin{align}
		\mathbf{h}_{\theta_k}(\tau) = e^{j\phi_k} \mathbf{\sigma}_k \delta(\tau),
	\end{align}
	where \utannotation{$\delta(\tau)$ is the Dirac distribution, }$\phi_k$ is the phase shift due to the propagation delay to the array, and $\mathbf{\sigma}_k$ is the steering vector to user $k$.  The phase shift is assumed to be uniformly distributed over $[0,2\pi]$.  The $m$\mbox{-}th element of the steering vector, in the case of a linear array with uniform spacing, is given by $[\mathbf{\sigma}_k]_m = e^{j2 \pi m \Delta \sin (\theta_k)/\lambda}$, where $\Delta$ is the distance between the antennas and $\lambda$ the wavelength of the signal carrier.  We study the case, where $\Delta = \lambda/2$, which is commonly regarded as the smallest interantenna distance that results in little coupling between antennas.
	
	In an environment with non-line-of sight, independent Rayleigh fading has proven to model the massive \MIMO channel well at symbol rate sampling \cite{gesbert2002outdoor}.  We assume the oversampled channel impulse response also to be i.i.d.\ Gaussian, i.e.\ each element in
	\begin{align}
		\operatorname{LP}\big\{\mathbf{H}(\tau)\big\}(\ell T / \kappa) \sim \mathcal{CN}(0, 1 / L),
	\end{align}
	where $\operatorname{LP}\{\cdot\}(t)$ is an ideal low-pass filter with cutoff frequency $\frac{\kappa}{2T}$ and where $L$ is the number of non-zero taps.  We study the case where $L = 15\kappa$, which corresponds to a maximum excess delay of 15 symbol periods.
	
	\subsection{Numerical Results}
	We define the in-band power, received adjacent-band power and maximum adjacent-band power as
	\begin{align}
	P_\text{ib}(\mathbf{\theta}) &\triangleq \int_{-B/2}^{B/2} S_\mathbf{\theta}(f) \mathrm{d}f,\\
	P_\text{ob}(\mathbf{\theta}) &\triangleq \max\Big\{\int_{-3B/2}^{-B/2}\!\! S_\mathbf{\theta}(f)\mathrm{d}f, \int_{B/2}^{3B/2}\!\! S_\mathbf{\theta}(f)\mathrm{d}f\Big\},\\
	P_\text{ob,max} &\triangleq \max \Big\{\!\int_{-3B/2}^{-B/2}\!\!\!\! S_\text{max}(f)\mathrm{d}f, \int_{B/2}^{3B/2}\!\!\!\! S_\text{max}(f)\mathrm{d}f \!\Big\}.
	\end{align}
	
\begin{figure}
		\centering
		\setlength{\figurewidth}{22em}
		\setlength{\figureheight}{25ex}
		\input{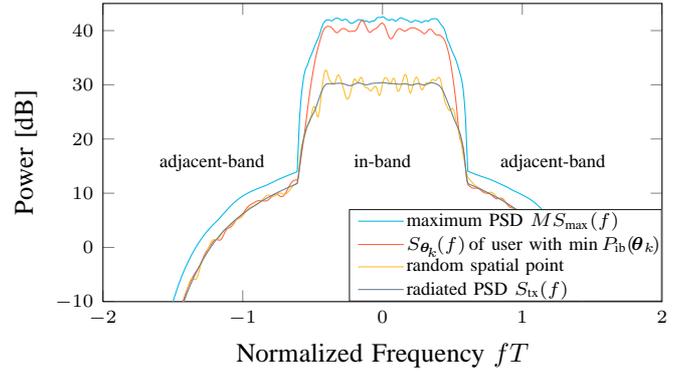}
		\caption{Power spectral densities for a system with 10~users and 100~antennas in a Rayleigh fading channel.}
		\label{fig:psds_K10_M100}
	\end{figure}

	The power spectral densities in Figure~\ref{fig:psds_K10_M100} are from a system with 100~base station antennas that serves 10~users over a realization of a frequency-selective Rayleigh fading channel.  Because of channel hardening, generating another channel does not change the general appearance of the curves.  By measuring the vertical distance between the transmitted power spectral density $ S_\text{tx}(f)$ (black) to the power spectral density $S_{\mathbf{\theta}_k}(f)$ received at the user with the smallest $P_\text{ib}(\mathbf{\theta}_k)$ (red), we see that the array gain\footnote{With 100 antennas, the maximum array gain is \unit[20]{dB}.  Here 10 users share this, so each user gets \unit[10]{dB} of array gain.} of the in-band power of even the weakest user is around \unit[10]{dB}.  Furthermore we see, when the maximum power spectral density $\Exp[\|\mathbf{\tilde{h}}(f)\|^2] S_\text{max}(f)$ (blue) is compared to the transmitted power spectral density $S_\text{tx}(f)$, that the worst-case out-of-band power has a much smaller array gain, around \unit[2]{dB}.  The received power spectral density $S_\mathbf{\theta}(f)$ at many random points $\mathbf{\theta}$ were generated, each with an independent Rayleigh fading channel vector.  All had the same general appearance as the one that is plotted in yellow.  The received power varies around the radiated power level and is well below the maximum power spectral density.  
	
\begin{figure}
		\centering
		\setlength{\figurewidth}{22em}
		\setlength{\figureheight}{25ex}
		\input{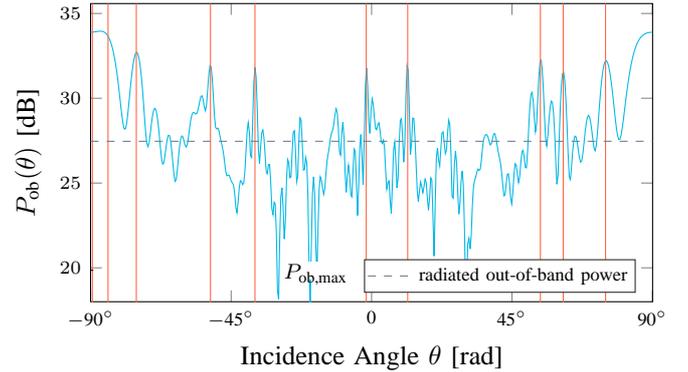}
		\caption{The adjacent-band power in different directions in a line-of-sight channel with 100 antennas and 10 users. The vertical lines indicate the directions of the users.}
		\label{fig:ob_radiation_pattern}
	\end{figure}

	In Figure~\ref{fig:ob_radiation_pattern}, the adjacent-band power $P_\text{ob}(\theta)$ of a line-of-sight system can be seen for different directions around the array.  From the peaks, it can be seen that the power out-of-band is beamformed in the directions of the served users.  The highest of these peaks, in this case, is \unit[4]{dB} above the transmitted adjacent-band power.  This is also how high the maximum adjacent-band power $P_\text{ob,max}$ (which upper bounds the adjacent-band power of any victim---not necessarily in line-of-sight) is above the transmitted adjacent-band power experienced by a victim.  The array gain of the worst-case adjacent-band power is thus slightly higher than in the Rayleigh fading case, but still significantly lower than the array gain seen in-band, which is \unit[10]{dB} (cannot be seen in the plot).  In between the served users, we see that the out-of-band power is approximately equal to or slightly lower than the radiated out-of-band power of $S_\text{tx}(f)$.  
	
\begin{figure}
		\centering
		\setlength{\figurewidth}{22.5em}
		\setlength{\figureheight}{25ex}
		\input{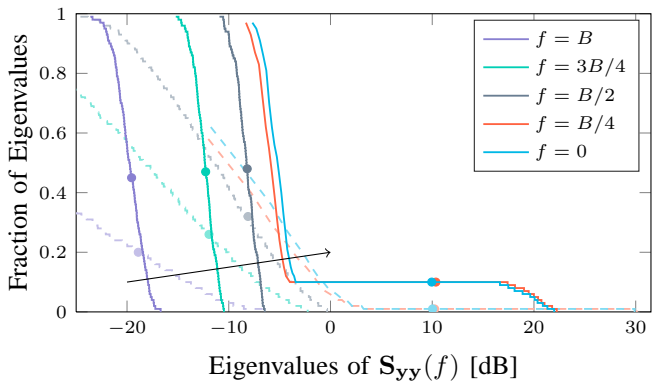}
		\caption{The complementary cumulative distribution of the eigenvalues of the correlation matrix $\mathbf{S}_{\mathbf{yy}}(f)$ at different frequencies $f$ for a Rayleigh fading channel with 100 antennas and 10 users (solid lines), and 1 user (dashed lines). The dot on each curve marks average eigenvalue $S_\text{tx}(f) / M$.}
		\label{fig:eigenvalue_dist}
	\end{figure}

	These observations can also be made by studying the eigenvalue distribution of the correlation matrix $\mathbf{S}_{\mathbf{yy}}(f)$ at different frequencies, see Figure~\ref{fig:eigenvalue_dist}, where a 100-antenna system that serves both 10 users and 1 user is studied for one realization of a Rayleigh fading channel.  We see that, for 10 users and frequencies $f < B/2$, 10 out of 100 eigenvalues are \unit[20]{dB} bigger than the rest.  These correspond to the directions of the users.  At out-of-band frequencies $f \geq B/2$ however, there are no eigenvalues significantly above the average, which is marked with by dot.  This means that, even in a worst-case scenario, a victim will not receive significantly more power out-of-band than on average.  
	
	In a single-user massive \MIMO system, the out-of-band radiation is distributed differently, see the dashed lines in Figure~\ref{fig:eigenvalue_dist}.  The signal out-of-band is more directive than in the multiuser case and has an array gain of approximately \unit[10]{dB} in the strongest direction.  This should be compared to the signal in-band, which has an array gain of \unit[20]{dB}.  We also see that \unit[20]{\%} of the eigenvalues are \unit[2]{dB} above the average at $f = \frac{B}{2}$, which means that the probability of an out-of-band radiation level that is higher than the average is significant.
		
\section{Conclusions}
	We have seen that massive \MIMO systems can operate with lower linearity requirements on the power amplifiers compared to conventional single-antenna systems without increasing the disturbance of communication in adjacent bands.  With the same linearity constraints and the same amount of radiated power, a victim that operates in an adjacent band will receive the same amount of disturbing out-of-band radiation from a single-antenna system as from a massive \MIMO system.  A massive \MIMO system can, however, due to its large array gain that grows with the number of antennas, lower its radiated power and still serve its users with the same quality of service as compared to the single-antenna system.  By lowering the radiated power, the amount of disturbing out-of-band power the victim receives from the massive \MIMO system decreases by the same amount.
	
	For specific realizations of the channel impulse response however, the small-scale fading of a victim might line up with the signal transmitted out-of-band and the victim then experiences much higher disturbing out-of-band radiation compared to the average.  Such a worst-case event can be a problem if (i) the fading is time-invariant or (ii) if it occurs often, which can only happen if the small-scale fading of the victim is correlated to the channels of the served users.  We have seen that the largest amount the out-of-band radiation received by a victim at a frequency $f$ can increase by in worst-case events is determined by the ratio $M S_\text{max}(f) / S_\text{tx}(f)$.  In multiuser scenarios, this ratio is small---\unit[2--4]{dB} with 100 antennas and 10 users.  In a single-user scenario however, this ratio can be much higher---in Rayleigh fading with 100 antennas, it is \unit[10]{dB}.  If coding can be done over multiple coherence intervals, however, worst-case events are not a problem since data lost during one coherence interval can be recovered.
	
	Further, we have seen that out-of-band radiation can be measured in space in terms of \MIMOACLR and that \MIMOACLR is the same as the per-antenna \ACLR measured at the base station.  To measure and constrain the radiated out-of-band power at the base station is thus sufficient to limit the average amount of power a victim in an adjacent band is disturbed by.
	
	\section*{Acknowledgement}
	The research leading to these results has received funding from the European Union Seventh Framework Programme under grant agreement number \textsc{ict}-619086 (\textsc{mammoet}) and the Swedish Research Council (Vetenskapsrådet).

	\bibliographystyle{IEEEtran}

	\bibliography{bib_forkort_namn,bibliografi}
\end{document}